\documentclass[letterpaper]{article} 
\usepackage{aaai2027}  
\usepackage[hyphens]{url}  
\usepackage{graphicx} 
\usepackage{natbib}  
\usepackage{caption} 
\usepackage{algorithm}
\usepackage{algorithmic}
\usepackage{subcaption}

\usepackage{newfloat}
\usepackage{listings}
\DeclareCaptionStyle{ruled}{labelfont=normalfont,labelsep=colon,strut=off} 
\floatstyle{ruled}
\newfloat{listing}{tb}{lst}{}
\floatname{listing}{Listing}

\usepackage{subcaption}
\usepackage{amsmath}
\usepackage{amsfonts}
\usepackage{booktabs}
\usepackage{multirow}
\usepackage{array}
\usepackage{tabularx}
\nocopyright

\usepackage{booktabs}

\newcommand{\eg}{\emph{e.g., }}

\title{Making Collaborative Signals Count: Graph-Aware Large Language Models for Sequential Recommendation}

\author{
Fenglin Yan\textsuperscript{\rm 1}\equalcontrib,
Bohao Wang\textsuperscript{\rm 1}\equalcontrib,
Jian Zhang\textsuperscript{\rm 2},
Yu Cui\textsuperscript{\rm 1},\\
Tongya Zheng\textsuperscript{\rm 1},
Ye Feng\textsuperscript{\rm 2},
Can Wang\textsuperscript{\rm 1},
Jiawei Chen\textsuperscript{\rm 1}\corresponding
}

\affiliations{
\textsuperscript{\rm 1}Zhejiang University, Hangzhou, China\\
\textsuperscript{\rm 2}University of Science and Technology of China, Anhui, China\\
\{fenglin.yan, bohao.wang, cuiyu23, wcan, sleepyhunt\}@zju.edu.cn\\
\{zhangjian833, yefengustc\}@mail.ustc.edu.cn,
doujiang\_zheng@163.com
}

\begin{document}

\maketitle

\begin{abstract}

Large language models (LLMs) have been widely adopted as backbones for recommender systems. However, their language-centric pretraining makes it difficult to capture collaborative signals implicit in user-item interactions, which are crucial for personalized recommendation. Existing methods either inject collaborative representations produced by external recommenders or model only intra-sequence dependencies, limiting their ability to exploit global collaborative patterns. To address this limitation, we propose GALLM, a graph-aware LLM framework for sequential recommendation. GALLM constructs a collaborative graph over text tokens and item tokens, and models three types of relations: Text--Text relations for preserving semantic dependencies, Item--Text relations for aligning item tokens with their textual descriptions, and Item--Item relations derived from global item co-occurrence patterns. These relations are transformed into lightweight learnable attention biases and incorporated into the LLM attention mechanism, enabling collaborative-aware token interactions without introducing an additional graph encoder. Experiments on four real-world benchmarks show that GALLM achieves the best performance among the compared baselines, improving over the strongest baseline by 9.76\% on average in HR@5. The code is available at https://github.com/Fenglin-miku/GALLM.
\end{abstract}

\section{Introduction}


Recommender systems (RS) aim to infer users' preferences from their historical interactions. A key to effective recommendation lies in capturing collaborative filtering (CF) signals from user--item interactions \citep{chen2020fast}. Different approaches exploit such signals from distinct perspectives: sequential recommenders capture temporal dependencies among items in chronologically ordered interaction sequences to model users' evolving preferences \citep{wang2019sequential, yu2019review, vaswani2017attention}, whereas graph-based recommenders propagate information over interaction graphs to capture global collaborative signals \citep{lightgcn, wang2019neural}.

In recent years, Large Language Models (LLMs) have demonstrated remarkable capabilities in semantic understanding, knowledge utilization, and reasoning \citep{achiam2023gpt, hao2026recreate}, motivating their increasing adoption in recommendation tasks \citep{tallrec, llara, slmrec}. A prominent line of research employs LLM as the backbone of RS \citep{collm, lost, text, msr}, typically converting users' interactions into natural language prompts and instructing the LLM to predict future interactions. By modeling fine-grained dependencies at the token level, LLMs can leverage semantic information to understand users' interests, representing a promising direction for advancing RS \citep{hao2026rethinking}.

\begin{figure}[t]
    \centering

    \includegraphics[width=0.55\columnwidth]{
        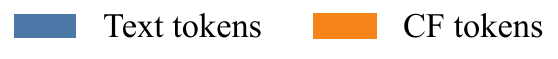
    }
    \begin{subfigure}[t]{0.49\columnwidth}
        \centering
        \includegraphics[width=\linewidth]{
            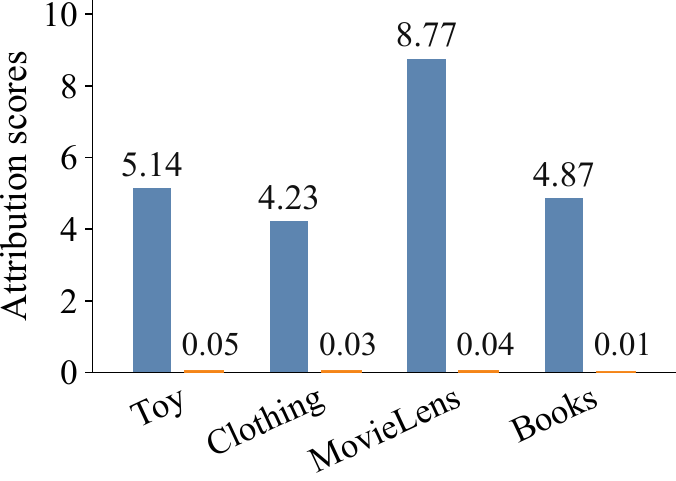
        }
        \caption{LLaRA}
        \label{fig:motivation_llara}
    \end{subfigure}
    \hfill
    \begin{subfigure}[t]{0.49\columnwidth}
        \centering
        \includegraphics[width=\linewidth]{
            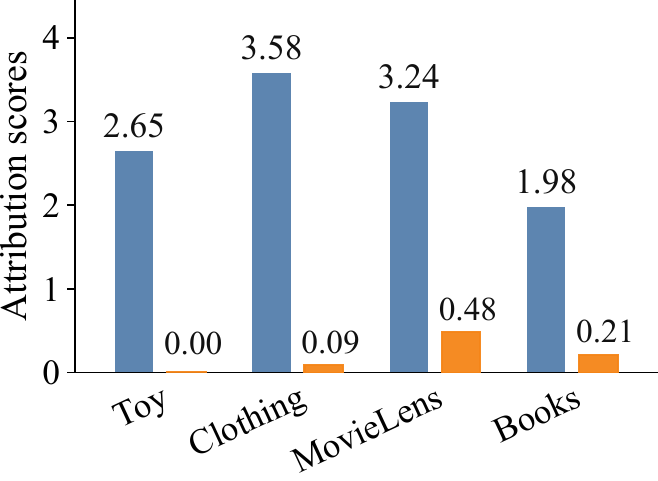
        }
        \caption{HeLLM}
        \label{fig:motivation_hellm}
    \end{subfigure}

    \caption{Attribution scores of Text tokens and CF tokens in LLaRA \cite{llara} and HeLLM \cite{hellm}. Higher scores indicate a greater influence on the generation of the target item.}
    \label{fig:motivation}
\end{figure}

\begin{figure*}[t]
    \centering
    \includegraphics[width=\textwidth]{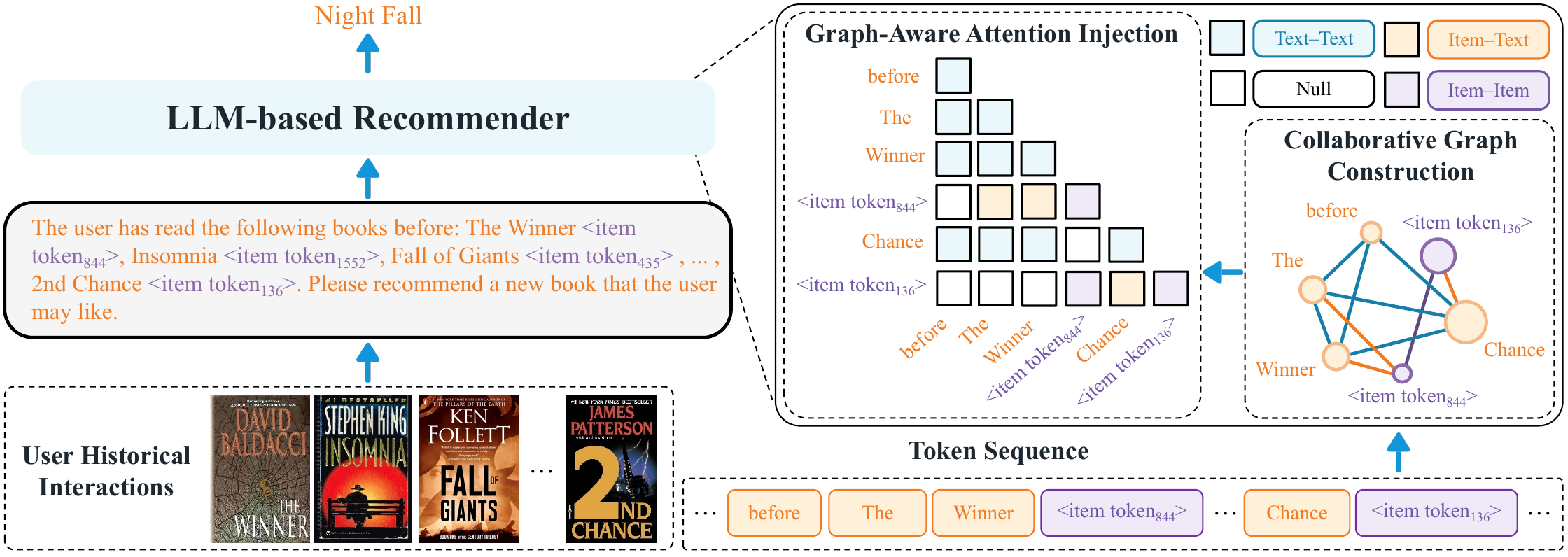}
    \caption{
    Overview of GALLM. Given a user's historical interactions, 
    we construct a collaborative graph containing Text--Text, Item--Text, 
    and Item--Item relations. The graph structures are encoded as attention 
    biases and incorporated into the LLM to enhance collaborative-aware 
    recommendation.
    }
    \label{fig:framework}
\end{figure*}

\begin{figure}[t]
    \centering
    \includegraphics[width=\linewidth]{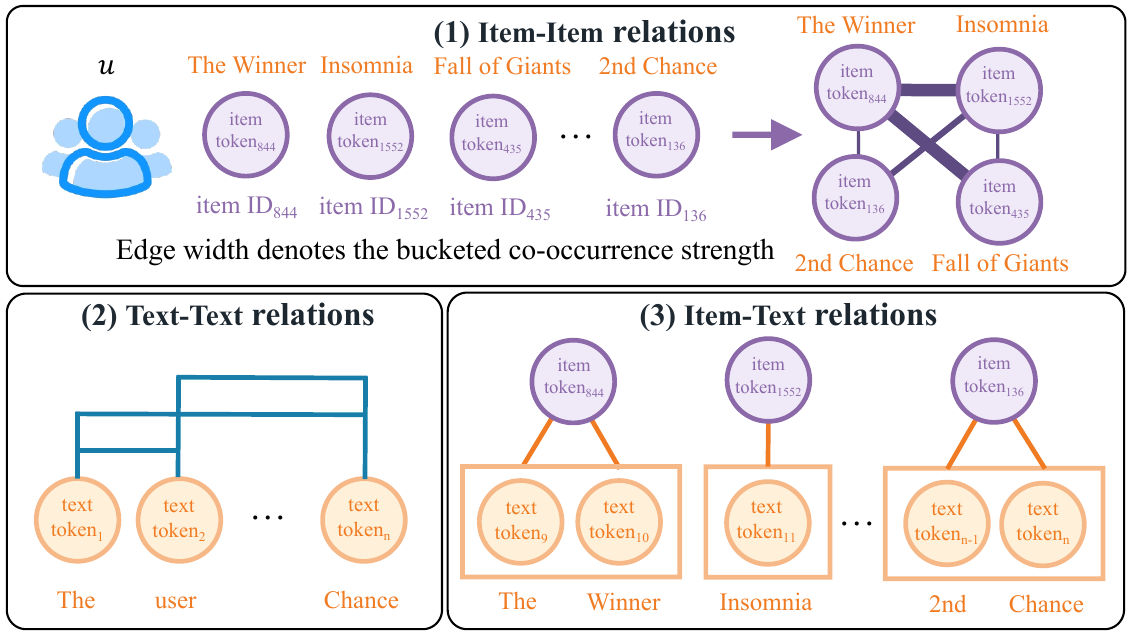}
    \caption{Illustration of the collaborative graph in GALLM, including Item--Item, Text--Text, and Item--Text relations.}
    \label{fig:heterogeneous_graph}
\end{figure}

Despite this promise, LLM-based RS remain limited in their capacity to model CF signals. This challenge arises from a mismatch between the token-level semantic dependencies modeled by LLMs and the item-level behavioral dependencies underlying CF signals. Recent studies have therefore explored enhancing CF modeling in LLM-based RS, broadly following two directions:
\begin{itemize}
    \item \textbf{Collaborative Injection via External Recommendation Models.}
    These methods employ auxiliary traditional recommendation models (e.g., SASRec \cite{sasrec}) to capture CF signals, which are subsequently incorporated into LLMs through collaborative embeddings \citep{llara}, predictive distributions \citep{tca4rec}, or model parameters \citep{cora}. However, their effectiveness is inherently limited by the capacity of the external models.
    Moreover, effectively integrating the extracted CF signals into LLMs remains challenging. A prevalent approach is to directly inject collaborative embeddings into the input of LLMs. However, such an approach often suffers from a misalignment between collaborative and textual representation spaces, making it difficult for LLMs to fully exploit the injected signals. Our empirical analysis in Figure~\ref{fig:motivation} further shows that collaborative embeddings contribute only marginally to model predictions and exert substantially less influence than textual tokens.
    \item \textbf{Collaborative Modeling via Attention Masking.}
    Some studies introduce specialized attention-masking mechanisms to enhance item-level dependency modeling within LLMs \cite{hatllm, iam}. Although these methods improve the modeling of sequential relationships among items, they primarily capture intra-sequence dependencies while overlooking crucial global collaborative dependencies across sequences.
\end{itemize}
These limitations motivate an important question: \textbf{\textit{How can LLMs effectively capture collaborative signals?}}

Graph-based methods provide a natural framework for modeling CF signals by organizing user--item interactions into a global interaction structure. Such graph structures expose collaborative information beyond individual interaction sequences and have been widely demonstrated to be effective in traditional recommendation systems \citep{lightgcn, wang2019neural, sasrec}. This naturally raises a question: can such graph structures be directly integrated into LLMs, enabling them to capture collaborative signals in a manner similar to traditional graph-based recommenders?

In this paper, we propose \textbf{Graph-Aware LLM-based Recommendation (GALLM)}, as illustrated in Figure~\ref{fig:framework}. Following prior work \citep{llara}, we represent each item using both textual tokens and a dedicated item token. GALLM further introduces explicit attention biases in Transformer layers to model the heterogeneous dependencies among different types of tokens. Specifically, we organize the resulting hybrid prompt as a token-level collaborative graph comprising three types of relations, as illustrated in Figure~\ref{fig:heterogeneous_graph}: (1) \textit{Item--Item relations}, which are derived from global item co-occurrence patterns to encode collaborative dependencies among items; (2) \textit{Item--Text relations}, which connect each item token to its associated textual tokens, thereby enhancing the model's ability to learn item-wise representations from the corresponding semantic context; and (3) \textit{Text--Text relations}, which preserve the semantic dependencies among textual tokens.

Despite its simplicity, this design enables each Transformer layer to perform relation-aware token interactions over the collaborative graph. Through successive layers, the entire model can be treated as a base LLM equipped with an implicit multi-layer attention-based graph aggregation mechanism, allowing GALLM to jointly model semantic and collaborative signals without introducing an external graph encoder.
Moreover, GALLM introduces only a small number of trainable parameters with no structural modifications to the LLM backbone, making it efficient and compatible with existing LLM architectures. Extensive experiments on four public benchmarks demonstrate that GALLM consistently outperforms state-of-the-art methods, achieving an average improvement of 7.62\% on NDCG@5 across all datasets.

The contributions of this work are summarized as follows:
\begin{itemize}
    \item We identify the limitations of existing approaches to enhance CF modeling in LLM-based RS propose integrating graph information into LLMs to address these limitations.
    \item We propose GALLM, a graph-aware LLM framework with lightweight learnable attention biases that enables joint modeling of semantic and global collaborative information within a unified LLM.
    \item We conduct extensive experiments on four public benchmarks, demonstrating that GALLM consistently outperforms state-of-the-art methods with an average improvement of 9.76\% on HR@5.
\end{itemize}

\section{Preliminary}

\subsection{Sequential Recommendation}
Let $\mathcal{U}$ and $\mathcal{I}$ denote the sets of users and items, respectively. For each user $u \in \mathcal{U}$, the historical interaction sequence is represented as $\mathcal{S}_u=(i_1,i_2,\ldots,i_{n-1})$, where $i_k \in \mathcal{I}$ denotes the $k$-th item interacted with by user $u$. The goal of sequential recommendation is to predict the next item $i_n$ based on the observed sequence $\mathcal{S}_u$.
Based on user--item interactions, the global item co-occurrence between items $i$ and $j$ is defined as
\begin{equation}
    C(i,j)=\sum_{u\in\mathcal{U}}
    \mathbb{I}(i\in\mathcal{S}_u)
    \mathbb{I}(j\in\mathcal{S}_u).
\end{equation}

A larger $C(i,j)$ indicates that items $i$ and $j$ co-occur in more users' interaction histories, suggesting a stronger collaborative dependency between them.

\subsection{LLM-based Recommendation}
Following prior studies \citep{bigrec, tallrec, llara, lin2024rella, hatllm, tallrec}, we also focus on the sequential recommendation task.
LLM-based sequential recommendation employs an LLM as the recommender backbone. The items in $\mathcal{S}_u$ are represented by textual descriptions (\eg titles), which are serialized into a language prompt $x_u$. Given $x_u$, the LLM is instructed to autoregressively generate the textual description of the target item, denoted by $y=(y_1,y_2,\ldots,y_{|y|})$. At each generation step $t$, the next token is predicted according to $P_{\theta}(y_t \mid x_u,y_{<t})$ where $\theta$ denotes the model parameters and $y_{<t}$ represents the previously generated tokens. By operating over textual tokens, this token-level formulation enables the LLM to capture fine-grained semantic patterns from the textualized interaction sequence.

\subsection{Graph Transformer}
Graph Transformers extend self-attention to graph-structured data by explicitly incorporating structural relations between nodes~\citep{ying2021transformers}. Let $\mathcal{G}=(\mathcal{V},\mathcal{E})$ denote a graph, where $\mathcal{V}$ and $\mathcal{E}$ are the node and edge sets, respectively. For two nodes $v_i,v_j\in\mathcal{V}$, a relation function $\phi(v_i,v_j)$ characterizes their structural relationship, such as connectivity, edge type, graph distance, or relation strength.

Formally, given the node representations
$\mathbf{H}^{l}=\{\mathbf{h}_1^{l},\ldots,\mathbf{h}_N^{l}\}$
at the $l$-th layer, Graph Transformers augment the content-based compatibility between two nodes with a relation-specific structural bias. The resulting pre-softmax attention logit is defined as
\begin{equation}
\widetilde{a}_{ij}^{l}
=
\frac{
(\mathbf{h}_i^{l}\mathbf{W}_{Q}^{l})
(\mathbf{h}_j^{l}\mathbf{W}_{K}^{l})^{\top}
}{
\sqrt{d}
}
+
b_{\phi(v_i,v_j)}^{l},
\label{eq:gt_attention_logit}
\end{equation}
where $b_{\phi(v_i,v_j)}^{l}$ is a learnable scalar associated with the structural relation between $v_i$ and $v_j$. $\mathbf{W}_{Q}^{l}$ and $\mathbf{W}_{K}^{l}$ are learnable query and key projection matrices, respectively, and $d$ is the dimension of the projected representations. In this way, self-attention jointly considers representation similarity and graph structural relations when aggregating information across nodes.

\section{Methodology}

\subsection{Overview}

In this work, we introduce GALLM, a graph-aware large language model framework for sequential recommendation, as illustrated in Figure~\ref{fig:framework}. 
The key idea is to leverage a collaborative graph to explicitly encode global collaborative structures into LLMs, enabling the unified modeling of semantic knowledge and CF signals.
Specifically, GALLM consists of three main components. 
First, we construct a hybrid recommendation prompt by integrating textual information with item tokens, allowing the model to simultaneously leverage semantic knowledge and collaborative signals. 
Second, we represent the hybrid prompt as a collaborative graph, in which heterogeneous relations among text and item tokens are explicitly modeled and global item co-occurrence patterns are incorporated as complementary CF signals.
Third, we introduce a relation-aware attention mechanism that transforms graph relations into learnable biases and injects them into LLM layers, enabling the model to capture collaborative dependencies while preserving its pretrained language capabilities.

\subsection{Hybrid Prompt Construction}
Given a user's historical interaction sequence
$\mathcal{S}_u=(i_1,i_2,\ldots,i_L)$, GALLM constructs a hybrid prompt by augmenting the conventional textual prompt with a dedicated Item token for each historical item.
Specifically, for the $k$-th historical item $i_k$, its textual description is tokenized into \textbf{Text tokens}:
\begin{equation}
\mathbf{x}_k^T=
(x_{k,1}^T,x_{k,2}^T,\ldots,x_{k,m_k}^T),
\label{eq:text_tokens}
\end{equation}
where $m_k$ denotes the number of Text tokens associated with item $i_k$. Meanwhile, an additional \textbf{Item token} $x_k^I$ is introduced after the textual tokens of $i_k$ to represent its collaborative information. The hybrid prompt is constructed by concatenating the instruction {Text tokens} $\mathbf{x}_{ins}^{T}$ with the item-specific token groups:
\begin{equation}
\mathbf{x}
=
(
\mathbf{x}_{ins}^{T};
\mathbf{x}_1^T,x_1^I;
\mathbf{x}_2^T,x_2^I;
\ldots;
\mathbf{x}_L^T,x_L^I),
\label{eq:hybrid_prompt}
\end{equation}
where each Item token embedding is parameterized as
$e_{i_k}^{item}\in\mathbb{R}^{d}$ for item $i_k$.
\footnote{Item token embeddings can be initialized from pretrained recommendation models (e.g., SASRec) via a lightweight MLP projection, providing collaborative priors.}

\subsection{Collaborative Graph Construction}
GALLM organizes the tokens in each hybrid prompt into a collaborative graph $\mathcal{G}=(\mathcal{V},\mathcal{E})$. Its node set $\mathcal{V}=\mathcal{V}_{T}\cup\mathcal{V}_{I}$ consists of Text tokens and Item tokens, while its edge set $\mathcal{E}=\mathcal{E}_{TT}\cup\mathcal{E}_{IT}\cup\mathcal{E}_{II}$ comprises three types of relations. 
Specifically, $\mathcal{V}_{T}$ contains the instruction Text tokens and item-specific Text tokens:
\begin{equation}
\mathcal{V}_{T}
=
\{x_{\mathrm{ins}}^{T}\}
\cup
\bigcup_{k=1}^{L}\{x_{k,j}^{T}\mid 1\leq j\leq m_k\},
\label{eq:text_nodes}
\end{equation}
while $\mathcal{V}_{I}$ denotes the node set consisting of Item tokens:
\begin{equation}
\mathcal{V}_{I}
=
\{x_k^I\mid 1\leq k\leq L\}.
\label{eq:item_nodes}
\end{equation}

To enable the LLM to jointly model textual semantics and CF information, we construct heterogeneous relations between Text tokens and Item tokens. These relations are defined according to the functional roles of different token pairs: Item--Item relations introduce global CF signals, Text--Text relations preserve contextual semantics, and Item--Text relations align Item token with their textual descriptions. This role-aware design provides task-relevant structural priors while preserving the original semantic modeling capability of the LLM. The three relation types are illustrated in Figure~\ref{fig:heterogeneous_graph} and detailed below.

\textbf{(1) Item--Item Relations} encode global collaborative signals from user--item interactions. Items that frequently co-occur in users' interaction histories are likely to share stronger behavioral dependencies.

Specifically, for two Item tokens $x_k^I$ and $x_{k'}^I$ corresponding to items $i_k$ and $i_{k'}$, respectively, we characterize their relation using the global co-occurrence frequency $C(i_k,i_{k'})$. Since raw co-occurrence frequencies are sparse and highly skewed, we discretize them into a set of relation categories $\mathcal{R}$ using equal frequency binning:
\begin{equation}
r_{kk'}
=
\mathrm{Bucket}\left(C(i_k,i_{k'})\right),
\quad
r_{kk'}\in\mathcal{R},
\label{eq:item_relation}
\end{equation}
where $\mathrm{Bucket}(\cdot)$ maps the co-occurrence frequency into predefined relation categories, and $\mathcal{R}$ denotes the set of discrete relation types. A larger relation index indicates stronger collaborative dependency between the two items. The resulting Item--Item edge set is then represented as:
\begin{equation}
\mathcal{E}_{II}
=
\left\{
(x_k^I,x_{k'}^I,r_{kk'})
\mid
1\leq k, k'\leq L
\right\}.
\label{eq:item_item_edge}
\end{equation}

These Item--Item edges distinguish different levels of collaborative relevance among items, enabling GALLM to capture global collaborative dependencies beyond intra-sequence item relationships.






\textbf{(2) Text--Text Relations} preserve semantic interactions among Text tokens in the hybrid prompt, allowing the LLM to retain its pretrained semantic modeling capability. The Text--Text edge set is defined over Text token nodes in $\mathcal{V}_{T}$:
\begin{equation}
\mathcal{E}_{TT}
=
\left\{
(x,x')
\mid
x,x'\in\mathcal{V}_{T}
\right\}.
\label{eq:text_text_edge}
\end{equation}
By preserving these connections, the collaborative graph retains the textual dependencies acquired during LLM pretraining and maintains the model's semantic modeling capability.

\textbf{(3) Item--Text Relations} connect each Item token to its corresponding Text tokens.
Specifically, for the Item token $x_k^I$ and the Text tokens of item $i_k$, the Item--Text edge set is defined as:
\begin{equation}
\mathcal{E}_{IT}
=
\left\{
(x_k^I,x_{k,j}^{T})
\mid
1\leq k\leq L,\;
1\leq j\leq m_k
\right\}.
\label{eq:item_text_edge}
\end{equation}
These relations explicitly align each Item token with its item-specific textual context, facilitating the learning of coherent item-wise representations that integrate collaborative and semantic information.

Through the collaborative graph, GALLM explicitly organizes the hybrid prompt into a structured representation that captures semantic dependencies, collaborative signals, and their alignment, providing structural knowledge for the subsequent relation-aware attention mechanism.

\subsection{Graph-Aware Attention Injection}

To enable the LLM to leverage the structural information encoded in the collaborative graph, we introduce a graph-aware attention mechanism. Inspired by Graph Transformers \citep{ying2021transformers}, GALLM introduces a graph-aware attention mechanism that encodes relation-specific graph information as learnable attention biases and incorporates them directly into the LLM self-attention computation.

Specifically, for each Transformer layer $l$, given the hidden representations
$\mathbf{H}^{l}=\{\mathbf{h}_1^{l},\ldots,\mathbf{h}_m^{l}\}$ of the input tokens, where $m$ denotes the total number of Text and Item tokens in the hybrid prompt, GALLM modifies the attention logits by introducing a relation-specific bias:
\begin{equation}
\widetilde{a}_{ij}^{l}
=
\frac{
(\mathbf{h}_i^{l}\mathbf{W}_{Q}^{l})
(\mathbf{h}_j^{l}\mathbf{W}_{K}^{l})^{\top}
}{
\sqrt{d}
}
+
b_{\phi(x_i,x_j)}^{l},
\label{eq:gt_attention_logit}
\end{equation}
where $\mathbf{W}_{Q}^{l}$ and $\mathbf{W}_{K}^{l}$ denote the query and key projection matrices of the $l$-th Transformer layer. Here, $x_i,x_j\in\mathcal{V}$ denote two token nodes in the collaborative graph, $\phi(x_i,x_j)$ represents their relation type, and $b_{\phi(x_i,x_j)}^{l}$ is a learnable relation-specific bias. The attention distribution is then computed based on the enhanced attention logits.



We define three groups of relation biases according to the collaborative graph: $b_{TT}^{l}$ for Text--Text relations, $b_{IT}^{l}$ for Item--Text relations, and $b_{II,\mathcal{R}}^{l}$ for Item--Item relations with different collaborative strengths. For all token pairs $(i,j)$, the original causal mask is retained, and relation biases are applied only to causally visible positions.
\begin{equation}
b_{\phi(x_i,x_j)}^{l}
=
\begin{cases}
b_{TT}^{l},
& (x_i,x_j)\in\mathcal{E}_{TT},\\
b_{IT}^{l},
& (x_i,x_j)\in\mathcal{E}_{IT},\\
b_{II,\mathcal{R}}^{l},
& (x_i,x_j)\in\mathcal{E}_{II}, \mathcal{R}=r_{ij},\\
0,
& \text{otherwise}.
\end{cases}
\label{eq:relation_bias}
\end{equation}

By incorporating lightweight relation-specific biases into self-attention, each Transformer layer performs relation-aware information aggregation over the collaborative graph. As the layers are stacked, token representations are iteratively updated by aggregating information from structurally related nodes, forming an implicit multi-layer graph aggregation process within the LLM. This process allows collaborative structural information to propagate across token representations while jointly modeling semantic and collaborative signals.




\section{Experiments}
We aim to answer the following research questions:
\begin{itemize}
    \item \textbf{RQ1:} How does GALLM perform compared with state-of-the-art recommendation methods?
    \item \textbf{RQ2:} How do different components of GALLM contribute to the recommendation performance?
    \item \textbf{RQ3:} How does GALLM influence the attention patterns of the LLM?
\end{itemize}

\begin{table}[t]
\centering 
\setlength{\tabcolsep}{1mm}
\begin{tabular}{c|cccc}
\toprule
Dataset & Toys & Clothing & Book & MovieLens \\ 
\midrule
\#User & 19124 & 39230 & 16559 & 11230 \\
\#Item & 11758 & 22948 & 6344 & 10681 \\
\#Interaction & 165247 & 277534 & 151928 & 165126 \\
Density & 0.0735\% & 0.0308\% & 0.1446\% & 0.0914\% \\
\bottomrule
\end{tabular}
\caption{Statistics of the datasets.}
\label{tab:dataset_statistics}
\end{table}

\begin{table*}[!ht]
\centering

{
\small
\setlength{\tabcolsep}{1.2pt}

\begin{tabularx}{\textwidth}{
>{\centering\arraybackslash}X|
>{\hspace{2pt}}c c c c<{\hspace{2pt}}|
>{\hspace{2pt}}c c c c<{\hspace{2pt}}|
>{\hspace{2pt}}c c c c<{\hspace{2pt}}|
>{\hspace{2pt}}c c c c<{\hspace{2pt}}
}
\toprule
\multirow[c]{2}{=}[-0.6ex]{\centering\textbf{Method}}
& \multicolumn{4}{c|}{\textbf{Toy}}
& \multicolumn{4}{c|}{\textbf{Clothing}}
& \multicolumn{4}{c|}{\textbf{Book}}
& \multicolumn{4}{c}{\textbf{MovieLens}} \\

\cmidrule(l{2pt}){2-17}

& H@5 & N@5 & H@10 & N@10
& H@5 & N@5 & H@10 & N@10
& H@5 & N@5 & H@10 & N@10
& H@5 & N@5 & H@10 & N@10 \\
\midrule

SASRec
& 0.0186 & 0.0096 & 0.0327 & 0.0142
& 0.0069 & 0.0033 & 0.0144 & 0.0057
& 0.0123 & 0.0074 & 0.0201 & 0.0099
& 0.0162 & 0.0094 & 0.0276 & 0.0130 \\

LightGCN
& 0.0027 & 0.0017 & 0.0042 & 0.0021
& 0.0008 & 0.0006 & 0.0014 & 0.0008
& 0.0027 & 0.0017 & 0.0050 & 0.0025
& 0.0064 & 0.0041 & 0.0122 & 0.0059 \\

\midrule

LLMRec
& 0.0104 & 0.0080 & 0.0148 & 0.0094
& 0.0039 & 0.0023 & 0.0060 & 0.0030
& 0.0118 & 0.0075 & 0.0198 & 0.0101
& 0.0219 & 0.0143
& 0.0548 & 0.0233 \\

CORONA
& 0.0142 & 0.0091 & 0.0264 & 0.0130
& \underline{0.0090} & \underline{0.0045}
& 0.0154 & 0.0070
& 0.0128 & 0.0072 & 0.0229 & 0.0105
& 0.0293 & 0.0173 & 0.0553 & 0.0256 \\

HeLLM
& 0.0191 & 0.0133 & 0.0325 & 0.0176
& 0.0080 & 0.0040 & 0.0172 & 0.0070
& \underline{0.0135} & \underline{0.0082}
& 0.0201 & 0.0104
& 0.0358 & 0.0222 & 0.0620 & 0.0306 \\

G2Rec
& 0.0211 & 0.0137 & 0.0346 & 0.0181
& 0.0082 & 0.0040
& \underline{0.0174} & \underline{0.0071}
& 0.0107 & 0.0060 & 0.0183 & 0.0085
& \underline{0.0376} & 0.0238 & 0.0632 & 0.0319 \\

\midrule

BIGRec
& 0.0217 & 0.0138 & 0.0340 & 0.0177
& 0.0078 & 0.0038 & 0.0160 & 0.0067
& 0.0123 & 0.0068 & 0.0228 & 0.0102
& 0.0302 & 0.0200 & 0.0542 & 0.0277 \\

\midrule

LLaRA
& 0.0227 & 0.0149 & \underline{0.0365} & 0.0186
& 0.0056 & 0.0032 & 0.0172 & 0.0069
& 0.0116 & 0.0071 & 0.0235 & 0.0109
& 0.0362 & 0.0234 & \underline{0.0644} & 0.0324 \\

HatLLM
& \underline{0.0232} & 0.0153 & 0.0346 & 0.0187
& 0.0070 & 0.0036 & 0.0164 & 0.0066
& 0.0132 & 0.0081 & 0.0224 & \underline{0.0117}
& 0.0372 & \underline{0.0241} & 0.0624 & \underline{0.0328} \\

TCA4Rec
& 0.0221 & \underline{0.0154}
& 0.0355 & \underline{0.0198}
& \underline{0.0090} & 0.0037 & 0.0166 & 0.0068
& 0.0116 & 0.0060 & \underline{0.0244} & 0.0102
& 0.0324 & 0.0195 & 0.0584 & 0.0279 \\

\midrule

\textbf{GALLM}
& \textbf{0.0252} & \textbf{0.0163}
& \textbf{0.0386} & \textbf{0.0206}
& \textbf{0.0096} & \textbf{0.0050}
& \textbf{0.0190} & \textbf{0.0073}
& \textbf{0.0157} & \textbf{0.0088}
& \textbf{0.0258} & \textbf{0.0120}
& \textbf{0.0404} & \textbf{0.0256}
& \textbf{0.0664} & \textbf{0.0340} \\

\midrule

\textbf{Impr.}
& 8.62\% & 5.84\% & 5.75\% & 4.04\%
& 6.67\% & 11.11\% & 9.20\% & 2.82\%
& 16.30\% & 7.32\% & 5.74\% & 2.56\%
& 7.44\% & 6.22\% & 3.11\% & 3.66\% \\

\bottomrule
\end{tabularx}
}

\caption{Performance comparison on four real-world datasets. The best and second-best results are boldfaced and underlined, respectively. ``H'' and ``N'' denote HR and NDCG. ``Impr.'' denotes GALLM's improvement over the strongest baseline.}
\label{tab:main_results}
\end{table*}

\subsection{Experimental Settings}
\textbf{Datasets.}
Several widely used datasets, including \textit{Amazon Toys and Games, Amazon Clothing, Amazon Books},\footnote{\url{https://nijianmo.github.io/amazon/index.html}} and \textit{MovieLens-10M}\footnote{\url{https://grouplens.org/datasets/movielens/10m/}} are employed in our experiments. The statistics of the processed datasets are summarized in Table~\ref{tab:dataset_statistics}.
For a fair comparison, we follow the preprocessing protocol of prior work~\citep{msl} and chronologically split each processed sequence into training, validation, and test sets with an 8:1:1 ratio. Further details are provided in Appendix~A.1.

\noindent \textbf{Baselines.} The baseline methods are grouped into several distinct categories, details are provided in Appendix~A.2.:
\begin{itemize}
\item \textbf{Traditional recommenders:} 
SASRec (ICDM'18) \citep{sasrec} and LightGCN (SIGIR'20) \citep{lightgcn} are included as representative conventional recommendation methods.
\item \textbf{LLM-based recommenders:}
BIGRec (TORS'25) \citep{bigrec} is included as a representative LLM-based RS that directly employs LLMs as recommendation backbones without explicitly incorporating CF signals.
\item \textbf{Graph-enhanced LLMs for recommendation:} 
LLMRec (WSDM'24) \citep{llmrec}, CORONA (SIGIR'25) \citep{corona}, HeLLM (arXiv'25) \citep{hellm}, and G2Rec (arXiv'26) \citep{g2rec} leverage graph to enhance LLM-driven RS.
\item \textbf{CF-enhanced LLMs for recommendation:}
LLaRA (SIGIR'24) \citep{llara}, HatLLM (arXiv'25) \citep{hatllm}, and TCA4Rec (WWW'26) \citep{tca4rec} are designed to enhance the capability of LLMs to capture CF signals.
\end{itemize}

\noindent \textbf{Implementation Details.}
Following previous work \citep{msl}, we evaluate recommendation performance using HR@$K$ and NDCG@$K$, where $K\in\{5,10\}$. We implement all baselines using their publicly available source codes and follow the settings recommended in the original papers for fair comparisons. For methods involving LLMs, we use LLaMA-3.2-3B \citep{grattafiori2024llama} as the backbone. Detailed implementation settings and hyperparameter configurations are provided in Appendix A.3.

\subsection{Performance Comparison (RQ1)}

Table~\ref{tab:main_results} reports the overall performance on four real-world datasets. We make the following observations.

\noindent\textbf{Overall performance.}
GALLM achieves the best results across all datasets and metrics, outperforming the strongest baseline by 9.76\% on average and by up to 16.30\% in HR@5. This demonstrates the effectiveness of encoding collaborative relations into LLMs.

\noindent\textbf{Comparison with traditional and graph-enhanced recommenders.}
Traditional recommenders SASRec and LightGCN mainly learn collaborative patterns from interaction data but lack LLM-based semantic understanding. Graph-enhanced methods LLMRec, CORONA, HeLLM, and G2Rec introduce graph information into recommendation, but often rely on external graph modules or graph-derived representations, limiting the direct integration of collaborative structures into LLM modeling.

\noindent\textbf{Comparison with LLM-based recommenders and CF-enhanced methods.}
BIGRec mainly leverages LLM semantic understanding without explicitly modeling collaborative signals. Existing collaborative-enhanced methods improve this limitation through external signal injection or attention-based mechanisms. However, injection methods LLaRA and TCA4Rec suffer from the difficulty of aligning CF representations with LLM semantic spaces, while attention-based methods HatLLM mainly capture local item dependencies within sequences and overlook global collaborative relations.

\noindent\textbf{Generalization across backbone scales.}
Table~\ref{tab:backbone_scalability} further compares GALLM with representative LLM-based methods using LLaMA3 backbones of different parameter scales. GALLM consistently outperforms BIGRec and LLaRA on both Toy and Clothing with the 1B, 3B, and 8B backbones. This consistent improvement demonstrates that the proposed graph-aware mechanism is effective across LLMs with different model capacities.

\begin{table}[t]
\centering
\small
{
\setlength{\tabcolsep}{4.8pt}
\begin{tabular}{c|c|cc|cc}
\toprule
\multirow[c]{2}{*}[-0.6ex]{\textbf{Backbone}}
& \multirow[c]{2}{*}[-0.6ex]{\textbf{Method}}
& \multicolumn{2}{c|}{\textbf{Toy}}
& \multicolumn{2}{c}{\textbf{Clothing}} \\

\cmidrule(l{2pt}){3-6}

& & H@10 & N@10 & H@10 & N@10 \\
\midrule

\multirow{3}{*}{LLaMA3-1B}
& BIGRec  & 0.0267 & 0.0129 & 0.0134 & 0.0053 \\
& LLaRA   & 0.0298 & 0.0135 & 0.0132 & 0.0054 \\
& \textbf{GALLM}
& \textbf{0.0315} & \textbf{0.0159}
& \textbf{0.0152} & \textbf{0.0063} \\

\midrule

\multirow{3}{*}{LLaMA3-3B}
& BIGRec  & 0.0340 & 0.0177 & 0.0160 & 0.0067 \\
& LLaRA   & 0.0365 & 0.0186 & 0.0172 & 0.0069 \\
& \textbf{GALLM}
& \textbf{0.0386} & \textbf{0.0206}
& \textbf{0.0190} & \textbf{0.0073} \\

\midrule

\multirow{3}{*}{LLaMA3-8B}
& BIGRec  & 0.0384 & 0.0214 & 0.0174 & 0.0078 \\
& LLaRA   & 0.0419 & 0.0224 & 0.0186 & 0.0076 \\
& \textbf{GALLM}
& \textbf{0.0461} & \textbf{0.0244}
& \textbf{0.0198} & \textbf{0.0081} \\

\bottomrule
\end{tabular}
}
\caption{Performance comparison using LLaMA3 backbones of different parameter scales.}
\label{tab:backbone_scalability}
\end{table}

\subsection{Ablation Study (RQ2)}












To investigate the contribution of different relations in the collaborative graph, we construct three variants of GALLM by removing Text--Text, Item--Text, and Item--Item relations, denoted as \textit{w/o T--T}, \textit{w/o I--T}, and \textit{w/o I--I}. We also compare with LLaRA, which only injects collaborative item embeddings without explicitly modeling graph relations.

As shown in Table~\ref{tab:ablation_relations}, removing any relation degrades performance, demonstrating that the three relations provide complementary information. Specifically, removing Text--Text relations causes performance drops on several datasets, indicating that preserving semantic dependencies among Text tokens is important for maintaining the LLM's semantic modeling capability. Removing Item--Text relations leads to degradation by breaking the alignment between collaborative Item tokens and their corresponding textual representations, limiting the interaction between semantic and collaborative information. Removing Item--Item relations consistently harms performance, verifying that global item co-occurrence relations provide essential global collaborative signals beyond directly observed interactions. Moreover, GALLM consistently outperforms LLaRA, showing that explicitly modeling heterogeneous relations within LLM attention is more effective than only injecting collaborative embeddings. Overall, the complete GALLM benefits from jointly capturing semantic dependencies, semantic--collaborative alignment, and collaborative structures.

\begin{figure}[t]
    \centering

    \includegraphics[width=0.58\columnwidth]{
        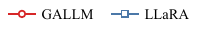
    }
    \vspace{1pt}

    \begin{subfigure}[t]{0.49\columnwidth}
        \centering
        \includegraphics[width=\linewidth]{
            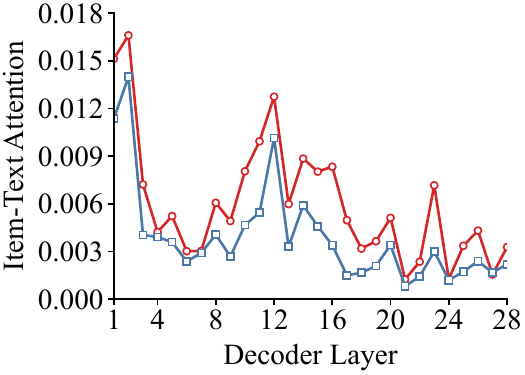
        }
        \caption{Toy}
        \label{fig:item_text_attention_toy}
    \end{subfigure}
    \hfill
    \begin{subfigure}[t]{0.49\columnwidth}
        \centering
        \includegraphics[width=\linewidth]{
            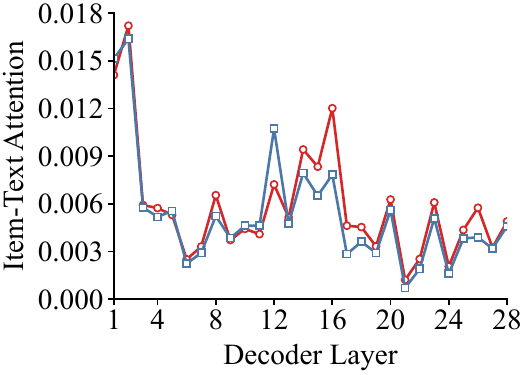
        }
        \caption{Book}
        \label{fig:item_text_attention_book}
    \end{subfigure}

    \caption{Mean Item--Text attention weights of GALLM and LLaRA on the Toy and Book datasets.}
    \label{fig:item_text_attention}
\end{figure}

\begin{table}[t]
\centering
\small
\setlength{\tabcolsep}{1.3pt}
\renewcommand{\arraystretch}{0.88}

\begin{tabular}{c|c|ccccc}
\toprule
\textbf{Dataset}
& \textbf{Metric}
& \textbf{GALLM}
& \textbf{w/o T--T}
& \textbf{w/o I--T}
& \textbf{w/o I--I}
& \textbf{LLaRA} \\
\midrule

\multirow{4}{*}{Toy}
& H@5  & \textbf{0.0252} & 0.0246 & 0.0246 & 0.0244 & 0.0227 \\
& N@5  & \textbf{0.0163} & 0.0153 & 0.0162 & 0.0157 & 0.0149 \\
& H@10 & \textbf{0.0386} & 0.0353 & 0.0367 & 0.0375 & 0.0365 \\
& N@10 & \textbf{0.0206} & 0.0187 & 0.0191 & 0.0200 & 0.0186 \\
\midrule

\multirow{4}{*}{Clothing}
& H@5  & \textbf{0.0096} & 0.0068 & 0.0058 & 0.0064 & 0.0056 \\
& N@5  & \textbf{0.0050} & 0.0037 & 0.0032 & 0.0034 & 0.0032 \\
& H@10 & \textbf{0.0190} & 0.0170 & 0.0172 & 0.0172 & 0.0172 \\
& N@10 & \textbf{0.0073} & 0.0070 & 0.0069 & 0.0068 & 0.0069 \\
\midrule

\multirow{4}{*}{Book}
& H@5  & \textbf{0.0157} & 0.0153 & 0.0148 & 0.0151 & 0.0116 \\
& N@5  & \textbf{0.0088} & 0.0085 & 0.0083 & 0.0080 & 0.0071 \\
& H@10 & \textbf{0.0258} & 0.0251 & 0.0257 & 0.0247 & 0.0235 \\
& N@10 & \textbf{0.0120} & 0.0117 & 0.0119 & 0.0115 & 0.0109 \\
\midrule

\multirow{4}{*}{MovieLens}
& H@5  & \textbf{0.0404} & 0.0378 & 0.0376 & 0.0370 & 0.0362 \\
& N@5  & \textbf{0.0256} & 0.0250 & 0.0240 & 0.0238 & 0.0234 \\
& H@10 & \textbf{0.0664} & 0.0654 & 0.0640 & 0.0644 & 0.0644 \\
& N@10 & \textbf{0.0340} & 0.0338 & 0.0324 & 0.0326 & 0.0324 \\
\bottomrule
\end{tabular}

\caption{Effect of different relation types in GALLM.}
\label{tab:ablation_relations}
\end{table}

\subsection{Attention Pattern Analysis (RQ3)}

\begin{figure}[t]\centering\includegraphics[width=0.48\textwidth]{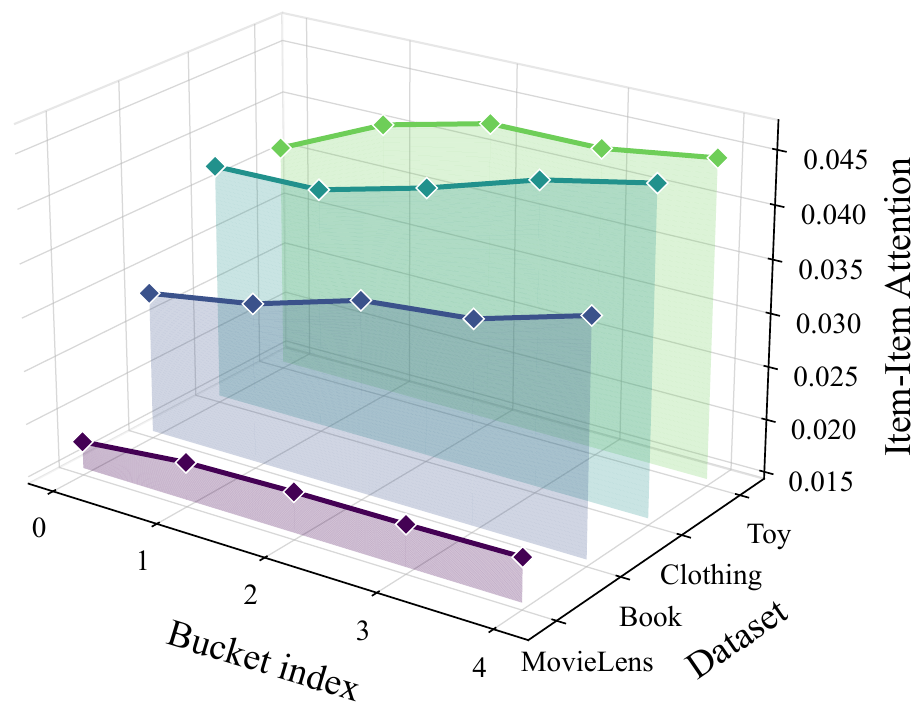}\caption{Mean Item--Item attention under different co-occurrence strengths across four datasets.}\label{fig:item_item_attention}\end{figure}

To investigate whether the proposed graph-aware attention mechanism effectively guides token interactions, we analyze post-softmax attention averaged over 28 decoder layers. For Item--Text attention, GALLM achieves higher alignment with textual descriptions than LLaRA across both datasets (Figures~\ref{fig:item_text_attention_toy} and~\ref{fig:item_text_attention_book}), increasing from 0.0039 to 0.0060 on Toy and from 0.0053 to 0.0059 on Book, with relative increases of 54.2\% and 10.0\%, respectively. For Item--Item attention, as shown in Fig.~\ref{fig:item_item_attention}, the average attention increases with co-occurrence strength across datasets, indicating that GALLM assigns higher weights to items with stronger collaborative relationships. These results verify that GALLM effectively reshapes LLM attention patterns by enhancing item--text semantic alignment and capturing item-level collaborative dependencies.

\section{Related Work}

\subsection{Sequential Recommendation}

Sequential recommendation predicts users' next interactions based on historical behaviors \citep{boka2024survey}. Early neural approaches employ recurrent or convolutional architectures, such as GRU4Rec \citep{hidasi2015session} and Caser \citep{tang2018personalized}. Recent self-attention-based models, including SASRec \citep{sasrec} and BERT4Rec \citep{sun2019bert4rec}, effectively capture sequential dependencies among historical items.

\subsection{LLMs for Recommendation}

Large language models have recently been widely explored in recommendation due to their strong semantic understanding and knowledge transfer capabilities \citep{tallrec, cui2022m6, geng2022recommendation, bigrec, a-llm, llara, collm, cora, corona, hellm, wang2026msr}. Recent studies have increasingly focused on incorporating collaborative signals into LLM-based recommendation, leading to two representative research directions.

\textbf{Graph-Enhanced LLMs for Recommendation.}
Recent studies have explored incorporating graph structures into LLM-based recommendation. One line of work leverages LLMs as auxiliary components to enhance graph-based recommenders, such as augmenting user--item graphs with semantic knowledge or improving graph retrieval with LLM-generated knowledge \citep{llmrec, ren2024representation, corona, nlcf}. Another line of work directly incorporates graph-derived collaborative information into LLMs, enabling LLMs to perform end-to-end recommendation. These methods introduce graph-based representations, higher-order interaction relationships, or hypergraph-enhanced features into LLM inputs or attention mechanisms \citep{guan2025enhancing, wang2025llm, hellm, g2rec}. However, existing approaches either treat LLMs as knowledge providers decoupled from the recommendation process or rely on shallow graph signal injection, limiting the integration between collaborative structures and LLM representations.

\textbf{CF-enhanced LLMs for Recommendation.} This paradigm directly employs pretrained LLMs as recommendation backbones and reformulates user interactions into textual prompts. To enhance collaborative modeling, existing methods mainly follow two directions. One line of work injects collaborative signals extracted from external recommendation models into LLMs. LLaRA \citep{llara}, CoLLM \citep{collm}, A-LLMRec \citep{a-llm}, and BinLLM \citep{text} incorporate CF embeddings or collaborative representations into LLM inputs through different alignment strategies. CoRA \citep{cora} and GraphLoRA \citep{graphlora} further explore parameter-level injection to integrate collaborative knowledge into LLMs, while TCA4Rec \citep{tca4rec} combines CF prediction distributions with LLM generation outputs. HatLLM \citep{hatllm} and IAM \citep{iam} introduce attention masking or item-aware attention mechanisms to capture item-level dependencies within user interaction sequences. Nevertheless, these methods either suffer from embedding incompatibility when injecting external collaborative signals, or capture only intra-sequence dependencies while missing global collaborative dependencies cross-user collaborative patterns inherent in interaction graphs.

\section{Conclusion}


In this paper, we reveal that existing LLM-based recommender systems face a key limitation in effectively capturing collaborative filtering signals. To address this issue, we propose \textbf{GALLM}, a graph-aware LLM framework that injects collaborative graph structures into LLMs through lightweight relation-aware attention biases. By jointly modeling heterogeneous relations among text and item tokens, GALLM enables the unified modeling of semantic information and global CF signals. Extensive experiments on four real-world datasets demonstrate the effectiveness of our method. In the future, we will explore more adaptive graph construction strategies for graph-aware LLM recommendation.

\bibliography{aaai2027}


\clearpage
\appendix
\setcounter{table}{6}

\section*{A. Experiment Details}

\subsection*{A.1 Dataset Details}

We conduct experiments on three Amazon recommendation datasets, including Amazon Toys and Games, Amazon Clothing, and Amazon Books, obtained from the Amazon Review Data (2018)\footnote{\url{https://nijianmo.github.io/amazon/index.html}}, as well as MovieLens-10M obtained from GroupLens\footnote{\url{https://grouplens.org/datasets/movielens/10m/}}.

For Amazon Toys and Games and Amazon Clothing, we first apply 5-core filtering. For Amazon Books and MovieLens-10M, we randomly retain 100,000 items and their associated interactions before preprocessing due to their large scale. Then, we apply a sliding window of length 11 to user interaction sequences longer than 11 interactions. The resulting sequences are sorted in ascending order by timestamp and split into training, validation, and test sets with an 8:1:1 ratio. This chronological splitting ensures that each test interaction occurs after all training interactions, thereby preventing information leakage.

\subsection*{A.2 Baseline Details}

This section briefly introduces the baseline methods according to the categories used in the main text.

\noindent\textbf{Traditional recommenders:}
\begin{itemize}
\item \textbf{SASRec (ICDM'18)} employs self-attention to model users' evolving preferences from their historical interaction sequences.

\item \textbf{LightGCN (SIGIR'20)} learns user and item representations by propagating collaborative information over the user--item interaction graph using simplified graph convolution.

\end{itemize}

\noindent\textbf{LLM-based recommenders:}
\begin{itemize}
\item \textbf{BIGRec (TORS'25)} adopts a bi-step grounding paradigm that first fine-tunes an LLM to generate meaningful item tokens and then maps the generated tokens to actual items.
\end{itemize}

\noindent\textbf{Graph-enhanced LLMs for recommendation:}
\begin{itemize}
\item \textbf{LLMRec (WSDM'24)} employs LLMs to augment user--item interaction graphs with additional edges, node attributes, and user preferences, thereby improving graph-based recommendation.

\item \textbf{CORONA (SIGIR'25)} uses LLM-guided preference and intent modeling to progressively retrieve relevant user--item subgraphs, followed by GNN-based collaborative recommendation.

\item \textbf{HeLLM (arXiv'25)} combines hypergraph-enhanced multimodal representations with sequential behavioral information to improve LLM-based recommendation.

\item \textbf{G2Rec (arXiv'26)} constructs a sparse item co-engagement graph and combines graph-derived user interest representations with supervised semantic tokenization for generative recommendation.

\end{itemize}

\noindent\textbf{CF-enhanced LLMs for recommendation:}
\begin{itemize}
\item \textbf{LLaRA (SIGIR'24)} introduces hybrid prompting to align collaborative item embeddings learned by conventional recommenders with textual item representations in LLMs.

\item \textbf{HatLLM (arXiv'25)} employs hierarchical attention masking to emphasize intra-item semantic dependencies in shallow layers and cross-item collaborative dependencies in deeper layers.

\item \textbf{TCA4Rec (WWW'26)} projects item-level CF logits into token-level distributions and aligns them with LLM supervision through a soft next-token prediction objective.

\end{itemize}

\subsection*{A.3 Evaluation and Implementation Details}

\noindent\textbf{Baseline settings.}
We implement all baselines using their publicly available source code and follow the settings recommended in the original papers whenever applicable. For the traditional recommenders, SASRec and LightGCN, we use the Adam optimizer with a learning rate of 0.001, an embedding dimension of 64, and a batch size of 256. For all methods involving LLMs, we uniformly adopt LLaMA-3.2-3B. Other method-specific components and training settings follow their official implementations.

\noindent\textbf{GALLM settings.}
We fine-tune LLaMA-3.2-3B using LoRA with a rank of 8, a scaling factor of $\alpha=16$, and a dropout rate of 0.05. Following LLaRA, the Item token embeddings are obtained from SASRec item representations and mapped into the LLM hidden space through a two-layer MLP. GALLM is optimized using Adam with a learning rate of 0.001 and a batch size of 64. The model is trained for 5 epochs using the standard causal language modeling loss.

The co-occurrence strengths of the Item--Item relations are discretized into five relation buckets indexed by
$\mathcal{R}\in\{0,1,2,3,4\}$.
We evaluate the model after each epoch and select the checkpoint achieving the highest NDCG@5 on the validation set. During inference, we use trie-constrained beam search with a beam size of 10, restricting the generated outputs to valid item candidates.

\noindent\textbf{Evaluation protocol.}
We evaluate all methods using HR@K and NDCG@K, where $K\in{5,10}$. No negative sampling is used during evaluation, and all methods retrieve or rank candidate items over the complete item space. All item co-occurrence statistics used to construct the Item--Item relations are computed exclusively from the training split to prevent information leakage. All experiments are implemented in PyTorch and conducted on four NVIDIA RTX 5090 GPUs. Each experiment is independently repeated three times, and the average results are reported. Additional implementation details are included in the supplementary code.

\end{document}